\documentclass[aps,pra,twocolumn,showpacs,superscriptaddress]{revtex4-1}
\usepackage[colorlinks ,linkcolor=blue,anchorcolor=blue,citecolor=blue,urlcolor=blue]{hyperref}
\usepackage{graphicx}
\usepackage{epstopdf}
\usepackage{soul, color, xcolor}
\usepackage{bm}
\usepackage{siunitx} 
\usepackage{amsmath,amssymb,amsfonts}
\usepackage{appendix}

\def\be{\begin{equation}} \def\ee{\end{equation}}
\def\bea{\begin{eqnarray}} \def\eea{\end{eqnarray}}

\begin{document}

\title{Magnetically and electrically tunable nonlinear Hall effect in $\mathcal {T}t$-symmetric antiferromagnets}

\author{Yueyue Ning}
\affiliation{Center for Quantum Transport and Thermal Energy Science, Institute of Physics Frontiers and Interdisciplinary Sciences, School of Physics and Technology, Nanjing Normal University, Nanjing 210023, China}

\author{Yongliang Song}
\affiliation{Center for Quantum Transport and Thermal Energy Science, Institute of Physics Frontiers and Interdisciplinary Sciences, School of Physics and Technology, Nanjing Normal University, Nanjing 210023, China}

\author{Yanrui Ding}
\affiliation{Center for Quantum Transport and Thermal Energy Science, Institute of Physics Frontiers and Interdisciplinary Sciences, School of Physics and Technology, Nanjing Normal University, Nanjing 210023, China}

\author{Zhichao Zhou}
\affiliation{Center for Quantum Transport and Thermal Energy Science, Institute of Physics Frontiers and Interdisciplinary Sciences, School of Physics and Technology, Nanjing Normal University, Nanjing 210023, China}

\author{Xiao Li} 
\email{lixiao@njnu.edu.cn}
\affiliation{Center for Quantum Transport and Thermal Energy Science, Institute of Physics Frontiers and Interdisciplinary Sciences, School of Physics and Technology, Nanjing Normal University, Nanjing 210023, China}

\begin{abstract} 
	
$\mathcal{T}t$-symmetric antiferromagnets can exhibit multiferroicity with coexisting ferroelectricity and magnetic order, and thus host multiple order-parameter configurations.
However, how to electrically read out these configurations is an issue meriting further investigation. 
Here, by combining tight-binding and first-principles calculations, we propose electrically and magnetically tunable nonlinear Hall transport in ferroelectric $\mathcal{T}t$-symmetric antiferromagnets.
Model calculations show that ferroelectric switching can reverse the sign of the Berry curvature dipole (BCD), whereas the Néel vector rotation modulates its magnitude, allowing both ferroelectric and magnetic order to be reflected in the nonlinear Hall signal. 
First-principles calculations of the VOCl$_2$ monolayer confirm these behaviors and predict a sizable BCD that gives rise to a potentially detectable Hall voltage. 
Our findings open up opportunities to integrate electrical and magnetic modulations of the nonlinear Hall effect within a single material, with potential applications in multi-state information encoding and readout in antiferromagnetic spintronics.

\end{abstract} 
\pacs{} 
\maketitle

{\color{blue}\textit{Introduction.}} 
Electrical nonlinear transport refers to high-order current responses to an applied electric field~\cite{Ryohei2017,Tokura2018,Suárez-Rodríguez2025,Du2021}, with diverse applications, e.g., in full-wave rectification~\cite{Ideue2017,Isobe2020,Zhang2017} and frequency conversion~\cite{He2021,Min2023}.
Especially for antiferromagnetic order, the nonlinear Hall effect can serve as a powerful probe of the magnetization direction, and subsequently facilitates the efficient information readout in antiferromagnetic spintronics~\cite{Shao2020,Wang2021,Liu2021,Wang2023}.
Antiferromagnets are known to possess several advantages over ferromagnets, such as the absence of stray fields and ultrafast dynamics~\cite{Jungwirth2016,Baltz2018,Němec2018,Kampfrath2011,Kamil2018}.
However, their magnetization directions have long been difficult to determine.
Recent studies have pointed out that nonlinear Hall transport allows for the efﬁcient detection of the N\'{e}el vector orientation of collinear antiferromagnets~\cite{Shao2020,Wang2021,Liu2021,Wang2023}.
For example, although the linear Hall response is forbidden in $\mathcal{PT}$-symmetric antiferromagnets (where $\mathcal{P}$ is the spatial inversion symmetry and $\mathcal{T}$ is the time-reversal symmetry), an intrinsic nonlinear Hall current driven by the quantum metric is generated and varies with the magnetization~\cite{Liu2021}, thereby enabling the electrical determination of the magnetization orientation.

Apart from the $\mathcal{PT}$-symmetric antiferromagnets, there is a large family of antiferromagnetic materials with the $\mathcal{T}t$ symmetry (where $t$ denotes a translation of half the unit cell)~\cite{Mong2010,Šmejkal2018}.
Unlike $\mathcal{PT}$-symmetric antiferromagnets that belong to nonpolar magnetic groups and exclude spontaneous ferroelectric polarization, $\mathcal{T}t$-symmetric antiferromagnets are compatible with ferroelectricity and likely to exhibit multiferroicity~\cite{Guo2026}.
As a result, this class of antiferromagnetic materials can host multiple order-parameter configurations with distinct combinations of magnetization directions and ferroelectric states.
An intriguing question arises as to whether multiple configuration states in $\mathcal{T}t$-symmetric antiferromagnets can be read out via nonlinear Hall effect.
If so, given that the ferroelectric state can be tuned by an electric field, dual electric and magnetic manipulation of the nonlinear Hall effect may thus be realized in these materials.
However, related studies remain largely unexplored, and merit investigation for multiple encoding and readout of information in antiferromagnets.

In this work, we investigate the nonlinear Hall effect arising from the Berry curvature dipole (BCD) in ferroelectric $\mathcal{T}t$-symmetric antiferromagnets, by combining tight-binding and first-principles calculations.   
Our model calculations of a two-dimensional rectangular lattice demonstrate that the BCD flips sign between two ferroelectric states, allowing polarization reversal to switch the direction of the nonlinear Hall current.
Meanwhile, the BCD is continuously modulated by rotating the N\'{e}el vector. 
Moreover, taking the multiferroic VOCl$_2$ monolayer as a representative material, first-principles calculations of the BCD support the findings from the model. 
The BCD near the conduction band edge of the VOCl$_2$ monolayer reaches a considerable value of $\sim$$1.2\times10^{-2}$ Å, suggesting the BCD-induced nonlinear Hall response may be experimentally observed in the candidate material.
The above numerical results reveal that the nonlinear Hall effect can be tuned by both electric and magnetic means within a single $\mathcal{T}t$-symmetric antiferromagnet, and the effect can be utilized for probing multiple order-parameter configurations.
This work enables simultaneous electric and magnetic control over nonlinear Hall transport in a single antiferromagnet, offering a new design approach for advanced electronic devices toward multi-state information encoding and readout.

\begin{figure}[htb]
\includegraphics[width=8.5 cm]{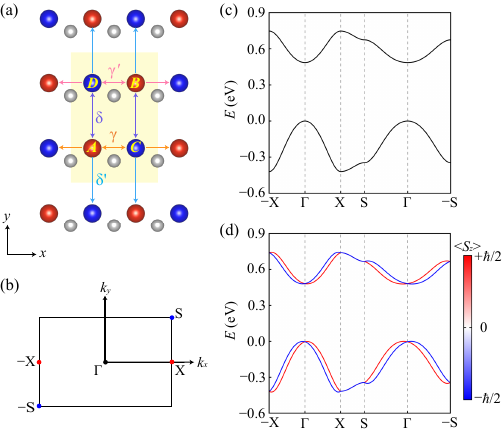}
\caption{ 
Schematic and band structures of a rectangular lattice with collinear antiferromagnetism. 
(a) A rectangular lattice with four sublattices, denoted as $A$-$D$. Red and blue balls represent magnetic atoms with opposite magnetic moments, while gray balls represent nonmagnetic atoms. Anisotropic NN and NNN hopping strengths are denoted by $\gamma$, $\gamma'$ and $\delta$, $\delta'$, respectively. The yellow shade represents a unit cell of the lattice. 
(b) The first Brillouin zone of the lattice. 
(c) and (d) Band structures of the lattice without and with the SOC, respectively. For both figures, the valence band maximum is set to zero energy, and the parameters $\varepsilon_{A,C}$ = $-\varepsilon_{B,D}$ = 1.0 eV, $\gamma$ = 0.3 eV, $\gamma'$ = 0.4 eV, $\delta$ = 0.1 eV, $\delta'$ = 0.2 eV, $J_{\mathrm{ex}}$ = $-$0.5 eV. Besides, $\lambda$ = $\lambda'$ = 0 eV for (c), while $\lambda$ = 0.03 eV and $\lambda'$ = 0.04 eV for (d). The out-of-plane spin projections are coded by color in (d).
}
\label{model_structure}
\end{figure}

{\color{blue}\textit{Model calculations.}}  We first consider a two-dimensional (2D) collinear antiferromagnetic model of a rectangular lattice, which is illustrated in Fig.~\ref{model_structure}(a). 
For the rectangular lattice, its unit cell consists of four sublattices, denoted as $A$-$D$. 
The nearest-neighbor (NN) and the next-nearest-neighbor (NNN) bonds between the sublattices are oriented along the $x$- and $y$-axes, respectively. 
Further considering the magnetic ordering, magnetic moments on $A$ and $B$ sublattices are ferromagnetically coupled, and they are antiparallel to those on $C$ and $D$ sublattices, forming a N\'{e}el-type antiferromagnetic configuration. 
Moreover, nonmagnetic atoms are asymmetrically arranged on one side of the NN $A$-$C$ and $B$-$D$ bonds, leading to local mirror symmetry breaking with respect to these bonds. 
As a result, this local structural asymmetry is expected to generate a potential gradient along the $y$-axis, pointing from nonmagnetic atoms to NN bonds. 
This potential gradient further induces the spin-orbit coupling (SOC)~\cite{Liu2011,Libor2020}, which will be discussed below.

Taking the crystal symmetry into account, the lattice preserves mirror symmetries, $\mathcal{M}_x$ and $\mathcal{M}_z$, with mirror planes perpendicular to the $x$- and $z$-axes, respectively, and a twofold rotational symmetry $\mathcal{C}_{2y}$ about the $y$-axis. 
In contrast, owing to the presence of the nonmagnetic atoms, the $\mathcal{M}_y$ symmetry is broken, which allows for a finite ferroelectric polarization along the $y$-axis. 
Assuming the ferroelectric polarization of the structure in Fig.~\ref{model_structure}(a) is along the $+y$ direction and denoted as the $+P$ state, the corresponding $-P$ state is related to the $+P$ state via the $\mathcal{M}_y$ operation.
Given the $\mathcal{C}_{2y}$ symmetry together with the $\mathcal {M}_y$ operation, the two states are also related by the $\mathcal{P}$ operation.
When the N\'{e}el-type antiferromagnetic order is further introduced into the lattice, the above crystal symmetries are possibly combined with the $\mathcal{T}$ operation, or broken, depending on the magnetization orientation. 
It is also noted that the antiferromagnetic structure has a combined $\mathcal{T}t$ symmetry, where $t$ is along the $x$-axis.
The $\mathcal{T}t$ symmetry distinguishes the antiferromagnetic model from the $\mathcal{PT}$-symmetric and altermagnetic ones~\cite{Libor2022}.

To investigate the electronic band structure and nonlinear Hall response of the $\mathcal{T}t$-symmetric antiferromagnetic model, we construct a tight-binding Hamiltonian. 
With the basis set of $s$ orbits of four magnetic sublattices, $\left\{\psi_{A, \uparrow}, \psi_{B, \uparrow}, \psi_{C, \uparrow}, \psi_{D, \uparrow}, \psi_{A, \downarrow}, \psi_{B, \downarrow}, \psi_{C, \downarrow},\psi_{D, \downarrow}\right\}$, the Hamiltonian reads 
\begin{equation}
	\begin{aligned}
		H &={} \sum_{i,\alpha}\varepsilon_i c_{i\alpha}^{\dagger}c_{i\alpha} + \sum_{\langle i,j\rangle,\alpha} \gamma_{ij}c_{i\alpha}^{\dagger}c_{j\alpha} 
		+ \sum_{\langle\!\langle i,j\rangle\!\rangle,\alpha} \delta_{ij}c_{i\alpha}^{\dagger}c_{j\alpha} \\ & - J_{\mathrm{ex}} \sum_{i,\alpha\beta} \left( \boldsymbol{\sigma}_{\alpha\beta}\cdot\textit{\textbf{S}}_i \right) c_{i\alpha}^{\dagger}c_{i\beta} 
		+  \sum_{\langle i,j\rangle,\alpha\beta}i\lambda_{ij}\nu_{ij}c_{i\alpha}^{\dagger}\sigma_{\alpha\beta}^{z}c_{j\beta}.
	\end{aligned}
	\label{Hamiltonian}
\end{equation}
Here, $\psi_{i, \alpha}$ is the wave function of the sublattice $i$ with spin $\alpha$ $(\alpha =~\uparrow, \downarrow)$, and $c_{i,\alpha}^{\dagger}\ (c_{i,\alpha})$ is the electron creation (annihilation) operator. 
The symbols $\left\langle i,j \right\rangle$ and $\left\langle \left\langle i,j \right\rangle \right\rangle $ represent NN and NNN sublattice pairs, respectively. 
The Hamiltonian consists of five terms, describing the on-site energy, NN hopping, NNN hopping, exchange interaction, and SOC, where $\varepsilon_i$, $\gamma_{ij}$, $\delta_{ij}$, $J_{\mathrm{ex}}$, and $\lambda_{ij}$ are strength parameters of these terms, respectively.
The symmetry breaking from the nonmagnetic atoms renders the strength parameters capable of taking two distinct values, as shown in Fig.~\ref{model_structure}(a).
The two distinct values for the parameter $\eta_{ij}$ ($\eta$ = $\gamma,~\delta,~\lambda$) are represented by $\eta$ and $\eta^\prime$.    
$\boldsymbol{\sigma} = (\sigma_x,\sigma_y,\sigma_z)$ are spin Pauli matrices.
$\boldsymbol{S}_{i}$ denotes the unit vector specifying the direction of the local magnetic moment on the sublattice $i$, and it is described by the polar angle, $\theta$, and the azimuthal angle, $\varphi$. 
For the collinear antiferromagnetic configuration in Fig.~\ref{model_structure}(a), $\boldsymbol{S}_{A} = \boldsymbol{S}_{B} = -\boldsymbol{S}_{C} = -\boldsymbol{S}_{D} = (\sin\theta\cos\varphi, \sin\theta\sin\varphi, \cos\theta)$. 
As discussed above, the SOC originates from the local potential gradient induced by the nonmagnetic atoms. 
For the horizontal NN hopping, the potential gradient along the $y$-axis gives rise to an effective magnetic field perpendicular to the lattice plane. 
In the SOC term, $\nu_{ij}$ = $\pm 1$, corresponding to the effective magnetic field along the $\pm z$ directions, respectively. 
Moreover, the parameters of the $-P$ state can be obtained from those of $+P$ state by the $\mathcal{M}_y$ operation.
Comparing the $\pm P$ states, two values of $\varepsilon_i$, $\gamma_{ij}$, and $\lambda_{ij}$ are interchanged, while those of $\delta_{ij}$ remain unchanged. 
Meanwhile, the reversal of the local potential gradient flips the out-of-plane effective magnetic field, leading to a sign change of $\nu_{ij}$.

According to the tight-binding Hamiltonian \ref{Hamiltonian}, we calculate the band structures of the antiferromagnetic model, with the magnetization along the $z$-axis. 
The band structures without and with SOC are presented in Figs.~\ref{model_structure}(c) and~\ref{model_structure}(d), respectively, where only the bands near the Fermi level are displayed. 
For the nonrelativistic band structure in Fig.~\ref{model_structure}(c), the bands are spin degenerate in the entire Brillouin zone. 
The degeneracy is protected by a spin group symmetry that combines the translation $t$ with a 2-fold rotation in the spin space around an axis perpendicular to the spins. 
The symmetry keeps the crystal wave vector $\textbf{\textit{k}}$ unchanged but reverses the spins, ensuring the spin degeneracy at each $\textbf{\textit{k}}$~\cite{Libor2022,Liu2022}.
Moreover, both the conduction band minimum and the valence band maximum are found to be located at $\Gamma$ point for the parameters considered here, forming a $\Gamma$ valley for both bands.

When the SOC is included in electronic structure calculations, it is found that the spin degeneracy is lifted in the relativistic band structure of Fig.~\ref{model_structure}(d), by the spin projection. 
In the neighborhood of $\Gamma$ valley, two spin-split bands acquire opposite momentum shifts away from the $\Gamma$ point, for both the conduction and valence bands.
As a result, there are opposite spins for electronic states at opposite crystal wave vectors that are related by the $\mathcal{T}$ operation, exhibiting the spin-momentum locking analogous to the spin-valley locking in MoS$_2$ monolayer~\cite{Xiao2012}. 
The spin-momentum locking is preserved by the $\mathcal{T}t$ symmetry, which serves as an effective $\mathcal{T}$ symmetry, while the real $\mathcal{T}$ symmetry is broken by the magnetic order.  
In contrast, the bands remain spin degenerate along the $X$-$S$ path, owing to the $\mathcal{M}_x$ symmetry.    

We then calculate the nonlinear Hall conductivities of the $\mathcal{T}t$-symmetric antiferromagnetic model.
The nonlinear current density, $J_a = \chi_{abc}E_bE_c$, describes quadratic response to an applied electric field, where $\chi_{abc}$ is the nonlinear conductivity tensor, and $E_b$, $E_c$ are the electric-field components, with $a$, $b$, and $c$ denoting Cartesian directions.
Since the intrinsic quantum-metric-induced nonlinear current is prohibited by the effective $\mathcal{T}$ symmetry~\cite{Wang2021,Liu2021}, we will focus on the nonlinear response from the BCD in the following. 
The BCD contributes to the second-order nonlinear conductivity tensor as given by the relation~\cite{Sodemann2015},
\begin{equation}
	\begin{aligned}
		\chi_{abc} = -\epsilon_{azc}\frac{e^3\tau}{2\hbar^2(1+i\omega\tau)}D_{b},
	\end{aligned}
	\label{chi}
\end{equation}
with the BCD for 2D materials,
\begin{equation}
	\begin{aligned}
		D_{b} = \int_{} \frac{d\textbf{\textit{k}}} {(2\pi)^2} d_{b} (\textbf{\textit{k}}) = \int_{} \frac{d\textbf{\textit{k}}} {(2\pi)^2} \sum_n f_{n\textbf{\textit{k}}} \frac{\partial \Omega^z_{n\textbf{\textit{k}}}} {\partial k_b},
	\end{aligned}
	\label{bcd}
\end{equation} 
and the Berry curvature along the $z$-axis~\cite{Xiao2010}, 
\begin{equation}
	\begin{aligned}
		\Omega^z_{n\textbf{\textit{k}}} = -2\hbar^2 \operatorname{Im} \sum_{m \neq n} \frac{ \langle \psi_{n\textbf{\textit{k}}} | v_x | \psi_{m\textbf{\textit{k}}} \rangle \langle \psi_{m\textbf{\textit{k}}} | v_y | \psi_{n\textbf{\textit{k}}} \rangle } { \left(E_{m\textbf{\textit{k}}} - E_{n\textbf{\textit{k}}}\right)^2 }.
	\end{aligned}
	\label{bc}
\end{equation}
Here, $e$, $\hbar$, $\tau$, $\epsilon_{azc}$, $\omega$ and $n$ are the unit charge, reduced Planck constant, relaxation time, Levi-Civita symbol, angular frequency of the applied electric field and band index, respectively. 
$d_{b}$(\textbf{\textit{k}}) and $f_{n\textbf{\textit{k}}}$ are the momentum-resolved BCD density and the equilibrium Fermi-Dirac distribution function, respectively.
$\psi_{n\textbf{\textit{k}}}$ is the wavefunction of Bloch electronic state and $E_{n\textbf{\textit{k}}}$ is corresponding energy.
$v_x$ and $v_y$ are velocity operators along the $x$- and $y$-axes, respectively.

\begin{figure}[htb]
	\includegraphics[width=8.5 cm]{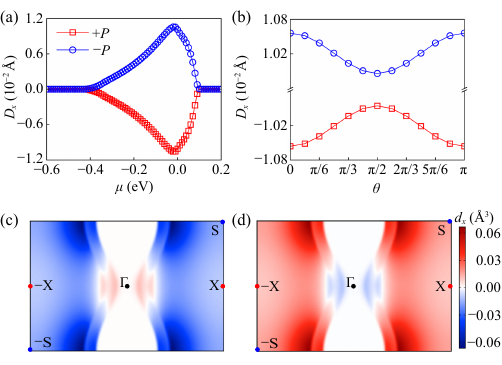}
	\caption{Berry curvature dipoles of the antiferromagnetic model for opposite ferroelectric states. 
	(a) Evolution of $D_{x}$ as a function of the chemical potential $\mu$, with the magnetization along the $z$-axis. 
	(b) Evolution of $D_{x}$ with the polar angle $\theta$. 
	(c) and (d) Momentum-resolved $d_{x}$($\textbf{\textit{k}}$) for $+P$ and $-P$ states, respectively, at $\theta = 0$. In (b)–(d), $\varphi=0$ and $\mu=-0.02$ eV.   
	}
	\label{model_bcd}
\end{figure}

Fig.~\ref{model_bcd}(a) shows the calculated BCD, $D_{x}$, as a function of the chemical potential $\mu$ for two opposite ferroelectric states of the antiferromagnetic model, with the magnetization along the $z$-axis. 
The results are presented for the energy range near the valence band edge, where the magnitude of $D_{x}$ is larger than that near the conduction band edge.
As $\mu$ moves away from the valence band maximum, the magnitude of $D_{x}$ in both $\pm P$ states first increases rapidly, reaches a maximum of $\sim$1.1$\times 10^{-2}$~\AA~at $\mu=-0.02~\mathrm{eV}$, and then decreases gradually toward zero. 
For each value of $\mu$, $D_x$ adopts the same magnitude but with opposite signs for the $\pm P$ states, indicating the ferroelectric polarization reversal can be used to switch the sign of the BCD. 
According to Eq.~\ref{chi}, a finite $D_x$ gives rise to a nonlinear conductivity tensor element, $\chi_{yxx}$, so that a BCD-induced nonlinear Hall current, $J_y=\chi_{yxx}E_x^2$, can be switched accordingly.
Besides, distinct from $D_x$, $D_y$ is calculated to be zero for both $\pm P$ states, which is ensured by the $\mathcal{M}_x$$\mathcal{T}$ symmetry.

The sign reversal of $D_x$ in Fig.~\ref{model_bcd}(a) can be understood through a symmetry analysis of the momentum-resolved BCD density, $d_x(\textit{\textbf{k}})$. Figs.~\ref{model_bcd}(c) and~\ref{model_bcd}(d) show $d_x(\textit{\textbf{k}})$ at $\mu=-0.02~\mathrm{eV}$ for $\pm P$ states, respectively.  
For both $\pm P$ states, it is seen that $d_x(\textit{\textbf{k}})$ has both equal magnitudes and the same sign at opposite wave vectors, which is ensured by the effective $\mathcal{T}$ symmetry, $\mathcal{T}t$, for arbitrary magnetization direction. 
Comparing Figs.~\ref{model_bcd}(c) and~\ref{model_bcd}(d), it is found that the values of $d_x(\textit{\textbf{k}})$ at the same $\textbf{\textit{k}}$ differ by the sign for opposite ferroelectric polarizations.
That is because the values at the same $\textbf{\textit{k}}$ are related by the $\mathcal{PT}$ operation. 
Given that the BCD density is odd under the $\mathcal{P}$ operation and possesses the aforementioned $\mathcal{T}$-even property, $d_x(\textit{\textbf{k}})$ experiences a sign reversal under the $\mathcal{PT}$ operation.
After integrating $d_x(\textit{\textbf{k}})$ over the Brillouin zone, $D_x$ is thus opposite for $\pm P$ states.
Besides, specific to the magnetization along the $z$-axis, the $\mathcal{M}_x$ symmetry preserved in both $\pm P$ states makes $d_x(\textit{\textbf{k}})$ symmetric about $k_x=0$. 
Together with $\mathcal{T}t$ symmetry, $d_x(\textit{\textbf{k}})$ is also symmetric about $k_y=0$, as shown in Fig.~\ref{model_bcd}(c) and 2(d).

We further investigate the effect of the magnetization direction on the BCD.
Fig.~\ref{model_bcd}(b) demonstrates $D_{x}$ as a function of the polar angle $\theta$ for the $\pm P$ states, with the azimuthal angle $\varphi=0$, which represents an out-of-plane rotation of the magnetization with respect to the 2D lattice plane.
For both $\pm P$ states, $D_{x}$ indeed varies with the magnetization direction, and exhibits a trigonometric behavior with a $\pi$ periodicity. 
The maximum and minimum of the magnitude of $D_x$ correspond to the magnetization along the $z$-axis ($\theta = 0$) and $x$-axis ($\theta=\pi/2$), respectively.
Moreover, $D_x$ is even with respect to $\theta=\pi/2$, as required by the $\mathcal{M}_z\mathcal{T}$ symmetry that connects the magnetization configurations with the angles $\theta$ and $\pi-\theta$ while keeping $D_{x}$ invariant. 
In addition, throughout all the magnetization directions, $D_x$ remains opposite for $\pm P$ states, which is enforced by their $\mathcal{P}$-paired structures, as mentioned above. 
These findings reveal that the magnitude of the BCD can be modulated by the magnetization direction, while its sign can be switched by reversing the ferroelectric polarization.
Therefore, the nonlinear Hall current is highly tunable by both magnetic and electric means, and it can be also used for distinguishing different ferroelectric orders and magnetization directions.

{\color{blue}\textit{Application to VOCl$_2$ monolayer.}}  Taking one step further from the model calculation, we then investigate nonlinear Hall responses in a candidate material, VOCl$_2$.
The bulk VOCl$_2$ is a layered van der Waals crystal that has been synthesized experimentally~\cite{Wang2025,H1997,Beiträge1981}, and its monolayer is thus expected to be readily obtained through mechanical exfoliation~\cite{Ai2019,Helmer2025}.
Moreover, the VOCl$_2$ monolayer is known to exhibit multiferroic properties~\cite{Ai2019,Mahajan2021,Tan2019}, with coexisting ferroelectric and collinear antiferromagnetic orders, which agrees with our model in Fig.~\ref{model_structure}.
By first-principles-based calculations, structural, electronic, and transport properties of the VOCl$_2$ monolayer are studied in the following.
The calculation details are provided in the Supplemental Material (SM).

\begin{figure}[htb]
	\includegraphics[width=8.5 cm]{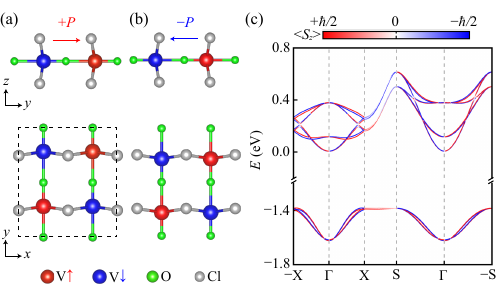}
	\caption{Crystal and band structures of the VOCl$_2$ monolayer. 
	(a) and (b) Side views (upper panels) and top views (lower panels) of $\pm P$ states, respectively. Red and blue balls represent V atoms with opposite magnetic moments, while green and gray balls denote O and Cl atoms, respectively. Arrows indicate the ferroelectric polarization directions. The unit cell is bounded by dashed lines. 
	(c) Band structure of the $+P$ state with the SOC, where the out-of-plane spin projections are coded by color. The conduction band minimum is set to zero energy. 
	}
	\label{vcl2o_structure}
\end{figure}

Figs.~\ref{vcl2o_structure}(a) and \ref{vcl2o_structure}(b) demonstrate side and top views of the multiferroic VOCl$_2$ monolayer in ferroelectric $+P$ and $-P$ states, respectively. 
The VOCl$_2$ monolayer exhibits a 2D rectangular lattice, with a crystal space group of $Pmm2$. 
It comprises three atomic layers, with a V-O layer sandwiched between two Cl layers. 
Each V atom is coordinated by two O atoms and four Cl atoms, forming a distorted octahedron.  
In the $+P$ state, V atoms are displaced along the $+y$ direction with respect to the coordinated anions, leading to a separation between positive and negative charge centers and consequently generating an in-plane spontaneous ferroelectric polarization along the $+y$ direction.
As for the $-P$ state, the displacement is reversed, giving rise to an opposite ferroelectric polarization. 
The atomic structure of the VOCl$_2$ monolayer shares the same crystal symmetries as the model discussed above.
The monolayer also has a N\'{e}el-type antiferromagnetic order, with magnetic moments contributed by four V atoms in a unit cell and neighboring magnetic moments aligned antiparallel along both the $x$- and $y$-axes.

For the VOCl$_2$ monolayer, its in-plane lattice constants are calculated to be 6.8~\AA~and 7.6~\AA~along the $x$- and $y$-axes, respectively.
The V-V distances along the two axes are $3.4$~\AA~and $3.8$~\AA, respectively, and the vertical distance between upper and lower Cl layers is $3.3$~\AA.
The magnitude of the local magnetic moment on each V atom is computed to be $1~\mu_{\mathrm B}$, and the total magnetic moment is zero for the unit cell.

Fig.~\ref{vcl2o_structure}(c) presents the first-principles band structure of the VOCl$_2$ monolayer in the ferroelectric $+P$ state, with considering the SOC and the magnetization along the $z$-axis.
In the band structure, the conduction band minimum is located at $\Gamma$ point, while the valence band maximum lies near X point.
The indirect band gap is computed to be 1.39 eV.
Moreover, according to the spin projection, the bands exhibit momentum-dependent spin splitting, which is consistent with the model calculations.

\begin{figure}[htb]
	\includegraphics[width=8.5 cm]{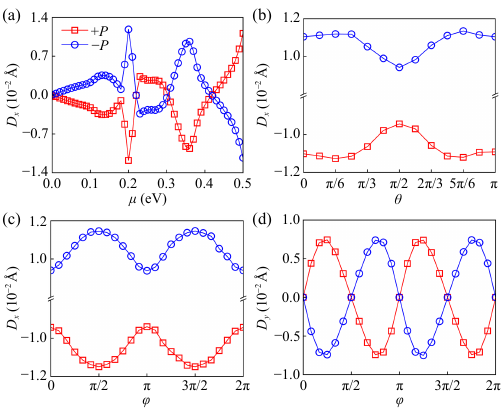}
	\caption{Berry curvature dipoles of the VOCl$_2$ monolayer for opposite ferroelectric states. 
	(a) Evolution of $D_{x}$ as a function of $\mu$, with the magnetization along the $z$-axis and the conduction band minimum set to zero energy. (b) Evolution of $D_{x}$ as a function of $\theta$ at $\varphi = 0$. (c) Evolution of $D_{x}$ as a function of $\varphi$ at $\theta = \pi/2$. (d) Evolution of $D_{y}$ as a function of $\varphi$ at $\theta = \pi/2$. In (b)-(d), $\mu = 0.2$ eV.
	}
	\label{vcl2o_bcd}
\end{figure}

The BCD is then calculated for the VOCl$_2$ monolayer, using the first-principles-based tight-binding Hamiltonian generated by the Wannier90 package~\cite{Pizzi2020}.
Fig.~\ref{vcl2o_bcd}(a) displays $D_x$ as a function of the chemical potential $\mu$ for $\pm P$ states, with the magnetization along the $z$-axis and the energy range near the conduction band edge.
It is seen that for an arbitrary $\mu$, $D_x$ has equal magnitudes but opposite signs for $\pm P$ states, which indicates a sign reversal of the BCD upon ferroelectric polarization switching and agrees with the model calculations.
Moreover, $D_x$ for $\pm P$ states varies nonmonotonically with $\mu$, accompanied by multiple positive and negative peaks.
For example, when $\mu$ lies 0.2 eV above the conduction band minimum, $D_x$ exhibits a positive (negative) peak for the $-P$ ($+P$) state, with a magnitude of 1.2$\times$$10^{-2}$ \AA.
The peak value is considerable and expected to be measured experimentally.
Under transverse open-circuit conditions, the BCD-induced nonlinear Hall current can induce a quadratic Hall voltage $V_y$, which is estimated to be 2.94 $\mu$V for the peak value of $D_x$ (see SM).
This voltage is comparable to those measured in 2D WTe$_2$~\cite{Ma2019,Kang2019}, suggesting that the BCD-induced nonlinear Hall response is likely to be experimentally detectable in the VOCl$_2$ monolayer through electron doping.

Figs.~\ref{vcl2o_bcd}(b) and \ref{vcl2o_bcd}(c) further demonstrate evolutions of $D_x$ with the magnetization rotations in the $x$-$z$ plane ($\varphi=0$) and $x$-$y$ plane ($\theta=\pi/2$), respectively, corresponding to out-of-plane and in-plane rotations with respect to the monolayer plane. 
In both cases, for any magnetization rotation, $D_x$ still has opposite values for the $\pm P$ states.
Moreover, $D_x$ varies with both angles $\theta$ and $\varphi$, exhibiting a period of $\pi$.
For the out-of-plane rotation in Fig.~\ref{vcl2o_bcd}(b), the magnitude of $D_x$ takes a relatively large value at $\theta=0$ ($z$-axis) and a minimum value at $\theta=\pi/2$ ($x$-axis).
As for the in-plane rotation in Fig.~\ref{vcl2o_bcd}(c), it has minimum and maximum values at $\varphi=0$ ($x$-axis) and $\varphi=\pi/2$ ($y$-axis), respectively.
The evolution of $D_x$ can be thus utilized to distinguish opposite ferroelectric polarizations and to differentiate high-symmetry magnetization directions. 
In addition, unlike the model calculations, $D_y$ is calculated to be nonvanishing for the VOCl$_2$ monolayer with the in-plane magnetization, as shown in Fig.~\ref{vcl2o_bcd}(d).
$D_y$ also has a period of $\pi$ and behaves like a function of $\sin2\varphi$.
For the out-of-plane rotation of the magnetization, $D_y$ remains zero owing to the preserved $\mathcal C_{2y}$ symmetry.

{\color{blue}\textit{Discussion and summary.}}  The VOCl$_2$ monolayer investigated above represents a prototypical 2D ferroelectric $\mathcal{T}t$-symmetric antiferromagnet exhibiting electrically and magnetically tunable nonlinear Hall transport. 
$\mathcal{T}t$-symmetric antiferromagnets with polar magnetic groups offer a broader pool of candidate materials to realize the highly tunable transport, as the polar magnetic groups allow spontaneous ferroelectric polarization compatible with the antiferromagnetic order. 
Our first-principles BCD calculations for bulk BaCoF$_4$, as presented in SM, provide a realistic material example in three dimensions.
Moreover, other classes of antiferromagnets, such as altermagnets, can also support ferroelectricity~\cite{Guo2026}.
The electric and magnetic tuning of their transport properties merits further investigation.

As for potential device applications, ferroelectric polarization can be switched by an electric field~\cite{Dawber2005}, while the Néel vector of antiferromagnets may be reoriented through external magnetic field or staggered spin-orbit torques~\cite{J2014,Wadley2016}. 
Selective combinations of these two order parameters in antiferromagnets could be used for multi-state information encoding. 
These states could then be read out electrically by distinguishable nonlinear Hall signals. 
These write and read schemes provide a promising route to multi-state information storage in $\mathcal{T}t$-symmetric antiferromagnets.

To conclude, a magnetically and electrically tunable nonlinear Hall effect is studied in ferroelectric $\mathcal{T}t$-symmetric antiferromagnets, using tight-binding and first-principles calculations. 
The numerical calculations show that ferroelectric switching reverses the BCD sign, while the Néel vector reorientation changes its magnitude and symmetry-allowed components. 
The VOCl$_2$ monolayer is identified as a representative material with sizable nonlinear Hall responses. 
These findings offer opportunities for highly tunable nonlinear Hall transport and for designing advanced electronic devices targeted at antiferromagnetic spintronics.

{\color{blue}\textit{Acknowledgments.}}  This work was supported by the National Natural Science Foundation of China (Nos. 12374044, 12004186, 11904173). 

\appendix

%

\providecommand{\noopsort}[1]{}

\end{document}